\documentclass[aps,prl,preprint,superscriptaddress]{revtex4-2}

\def\be{\begin{equation}}
\def\ee{\end{equation}}

\def\md{\mathrm{d}}

\usepackage{bm}

\def\Res#1{\mathop{\mathrm{Res}}_{#1}}
\def\tz{\zeta}
\usepackage{esint}
\usepackage{graphicx}

\begin{document}


\title{Importance of the continuous spectrum in the excitation of sheared surface gravity waves}


\author{J.R.~Carpenter}
\affiliation{Institute of Coastal Ocean Dynamics, Helmholtz-Zentrum Hereon}



\date{\today}

\begin{abstract}
The initial value problem is solved for the excitation of long surface gravity waves in a continuously sheared flow.  This reveals the presence of a continuous spectrum along side the standard normal modes of gravity wave propagation. An analytical similarity solution for the evolution of the free surface displacement from the continuous spectrum is found for the impulse response to surface excitation. It is demonstrated that the continuous spectrum contribution can be a significant fraction of the surface response, with the amplitude of the continuous spectrum exceeding that of the upstream gravity wave mode for Froude numbers of order unity.  The Landau damped mode description of the continuous spectrum is found to provide a link between methods using dispersion relations for phase speeds within the range of the velocity profile, and the variable-shear profiles that do not admit normal modes in this range.
\end{abstract}


\maketitle






Surface gravity waves, as found on the ocean surface, are a classic and familiar example of wave physics.  They are generally well understood and predicted by a dispersion relation that relates the (horizontal) wave phase speed, $c$, to other parameters controling wave propagation.  These include the wavenumber $k = 2\pi/\lambda$ with $\lambda$ the wave length, the depth of fluid $H$, or any mean current velocity.  Deriving this dependence relies on our ability to identify discrete ``modes'' that completely describe the surface response to an arbitrary excitation.  These modes can be classified as either up- or downstream propagating in the case of a flowing layer in a channel.  However, when surface gravity waves are excited on a fluid layer with a horizontal current that has a shear that varies in the vertical ($z$), described in general by a horizontal velocity profile $U(z)$, the application of dispersion relations derived from linear theory exhibit some counter-intuitive results that appear to contradict with well established results from flows without varying shear.  In particular, discrete wave modes do not exist with phase speeds within the range of $U$ \cite{burn1953}.  This is also found to be the case for internal waves in general stratified shear flows under certain conditions \cite{bain1995,prat2000}.  An understanding of wave propagation for speeds close to $U$ is particularly important for interpreting hydraulic phenomena, such as the presence of control points, where long wave signal propagation changes between bi- and uni-directional \cite{prat2000,wint2014}.


Here it is shown that in flows with a vertically varying shear, a dispersion relation is insufficient to describe the excitation of surface gravity waves, and the discrete modes must be complemented with a continuous spectrum that has different spatio-temporal properties.  The evolution of an initial surface displacement in space and time can not be described solely by the normal modes.  The importance of a continuous spectrum is well known in different areas of physics, such as in plasma physics \cite{cair1985}.  However, with some notable exceptions, that focus on an alternate mathematical description and instability  \cite{balm1999,balm1995}, the continuous spectrum is generally not considered in explaining gravity wave behavior in free surface flows \cite{pere1976}.  This is despite it being identified in shear flows many decades ago \cite{brig1970}, and its importance being recently recognized in the water surface initial value problem \cite{elli2014}.  Previous concepts from these disciplines, such as Landau damping \cite{brig1970}, are found to be useful in their application to surface gravity wave excitation and lead to a clearer understanding of linear wave theory in more realistic flows with vertically varying shear.  


In the present work I (i) identify and generalize the presence of the continuous spectrum in shallow water shear flows, (ii) find an analytical solution for the space-time behaviour of the continuous spectrum, and (iii) demonstrate that the continuous spectrum can constitute a large fraction of the total water surface response to forcing \textemdash up to a third in the $U(z)$ profiles analyzed thus far.  This demonstrates a potentially very different space-time evolution of the surface response to excitation than if only the normal modes are considered.

\textit{Derivation}.  The evolution in time ($t$) of linear perturbations to an inviscid, sheared, shallow layer of fluid flowing in the horizontal $x$-direction is governed by Rayleigh's equation \cite{burn1953}:
\be \label{eq:rayleigh}
(U - c)\hat{w}'' - U''\hat{w} = \hat{\phi}(z;k,c) .
\ee
Here we denote the fluid vertical velocity by $w(z;x,t)$, and have performed both Fourier and Laplace transforms via $\tilde{w}(z;k,t) = \mathcal{F}\{ w(z;x,t)\}$ and $\hat{w}(z;k,c) = \mathcal{L}\{ \tilde{w}(z;k,t) \}$, with the standard Laplace variable $s$ replaced by $c$ through $s = -ikc$.  Primes denote ordinary differentiation with respect to the vertical coordinate $z$, and $\hat{\phi}(z;k,c)$ represents either an initial condition or forcing of the fluid vorticity.  The undisturbed depth of flow, $H$, is considered small compared to the horizontal scales, thus making the shallow water, or long wave approximation, which neglects a term proportional to $kH$ in Rayleigh's equation \cite{burn1953}.  If finite water depths were considered, it would lead to more complicated solutions that exhibit dispersion effects of the various wavenumber components.

A free surface is coupled to the motion described by (\ref{eq:rayleigh}) through the upper boundary conditions.  These comprise the jump condition expressing continuity of stress
\be \label{eq:jump}
(U_s - c)\hat{w}'(H) - U'_s\hat{w}(H) - ikg\hat{\eta} = \hat{\Gamma}(k,c)
\ee
and the kinematic condition
\be \label{eq:kc}
ik(U_s - c)\hat{\eta} = \hat{w}(H) + \hat{S}(k,c) ,
\ee
where the `$s$' subscript denotes evaluation at the surface $z=H$, $g$ the gravitational acceleration, $\eta(x,t)$ the vertical displacement of the surface, and $\Gamma(x,t)$, $S(x,t)$ represent an initial condition or forcing of the surface vorticity and displacement, respectively.  By combining (\ref{eq:jump}) and (\ref{eq:kc}) we can form a single upper boundary condition on $w$ of 
\be
(U_s - c)^2 \hat{w}'(H) - [U'_s(U_s - c) + g]\hat{w}(H) = g\hat{S} 
\ee
which together with a no flow condition of $\hat{w}(0) = 0$ on the bottom boundary, completes a description of the problem.  Note that $\Gamma$ has been absorbed into $S$ to describe a single surface initial condition or forcing.

Solutions to the problem can be expressed in terms of the appropriate Green's function \cite{case1960} through
\be \label{eq:w_soln}
\hat{w}(z;k,c) = \int_0^H G(z,\zeta) \hat{\phi}(\zeta) \: \md \zeta - \frac{g\hat{S}(k,c)}{(U_s - c)}G(z,H) ,
\ee
where 
\be
G(z,\zeta) = \frac{U(z) - c}{\mathcal{D}(c)} \cdot \left\{ \begin{array}{c@{\:}l} 
\psi_<(\tz)[\psi_>(z) + g^{-1}] & \textrm{, $z > \tz$} \\\relax
\psi_<(z)[\psi_>(\tz) + g^{-1}]  & \textrm{, $ z < \tz$} \\   \end{array} \right. ,
\ee
with $\psi(z;c)$ given by
\be
\psi_<(z;c) = \int_0^z \frac{\md z'}{[U(z') - c]^2} \: \: ; \: \: \psi_>(z;c) = -\int_z^H \frac{\md z'}{[U(z') - c]^2} .
\ee
Of central importance in this solution is the dispersion function 
\be \label{eq:D}
\mathcal{D}(c) \equiv \psi_<(z;c) - \psi_>(z;c) - g^{-1}
 = \int_0^H \frac{\md z}{[U(z) - c]^2} - g^{-1}.
\ee

It has long been known that zeros of $\mathcal{D}(c)$ define the dispersion relation, and give the discrete phase speeds of the surface gravity wave modes \cite{burn1953,pere1976}.  However, it can be seen that the integral in $\mathcal{D}(c)$ is hypersingular (since the denominator vanishes with a power greater than unity, see \cite{ang2013}) and non-divergent for real $c$ within the range of $U(z)$.  This fact hints that it is not sufficient to consider only the discrete modes, and that an initial value approach is required, as originally identified in an analogous problem in plasma physics by Landau \cite{land1946}.

To simplify solutions, only forcing of the surface displacement [$\hat{\phi}(z) = 0$, and $S(x,t) \neq 0$] is considered further, since the solution in (\ref{eq:w_soln}) is a sum of the responses to the $\hat{\phi}$ and $\hat{S}$ forcing terms, and the $\hat{\phi}$ forcing has been considered in other shear flows \cite{farr1996}.  Then the transformed surface displacement can be written using (\ref{eq:kc}) as
\be
\hat{\eta}(k,c) = -\frac{\hat{S}(k,c)}{ikg(U_s - c) \mathcal{D}(c)} \equiv -\frac{\hat{S}}{ik} \mathcal{H}(c)
\ee 
upon defining the function $\mathcal{H}(c)$.  Inversion of the Laplace transform is given by
\be
\tilde{\eta}(k,t) = \mathcal{L}^{-1}\{ \hat{\eta}(k,c) \} = \frac{1}{2\pi i} \int_{\infty + i \gamma}^{-\infty + i\gamma} \mathcal{H}(c) \hat{S}(k,c) e^{-ikct} \: \md c ,
\ee
with $\gamma$ chosen sufficiently large to ensure that the integration path is above all singularities in the integrand.  Insight can be gained by considering the impulse response with $S(x,t) = \delta(x)\delta(t)$, or equivalently, $\hat{S}(k,c) = 1$, through which solutions to other forcings or initial conditions can be generated by convolution.  Then inversion of the Laplace transform can be accomplished by consideration of the singularities of $\mathcal{H}(c)$, where we can write
\be \label{eq:eta_kt}
\tilde{\eta}(k,t) = \sum_n \alpha_n e^{-ikc_n t} + \frac{1}{2\pi i}\int_B \mathcal{H}(c) e^{-ikct} \: \md c \equiv \tilde{\eta}_\mathrm{dis}(k,t) + \tilde{\eta}_\mathrm{cts}(k,t) .
\ee
The first term on the rhs ($\tilde{\eta}_\mathrm{dis}$) represents the isolated pole singularities, whose location in the complex $c$-plane is described by the dispersion relation $\mathcal{D}(c) = 0$, thus giving the contribution of the discrete modes.  The amplitudes of these modal solutions, $\alpha_n$, are determined by the residues of $\mathcal{H}$ at these singularites, i.e., $\alpha_n = \Res{c=c_n} \mathcal{H}(c)$.  The second term ($\tilde{\eta}_\mathrm{cts}$) represents the continuous spectrum, which will be described shortly.

\textit{Waves on an unsheared current}.  At this point, it is helpful to consider the simple reference case of shallow water waves on an unsheared current with constant velocity with depth, i.e., $U(z) = U_s$ for $0 \leq z \leq H$.  In the specific cases analyzed henceforth (unsheared and parabolic currents), we shall non-dimensionalize all quantities using the length scale $H$ and velocity scale $U_s$, and this will result in dependence on the Froude number $F \equiv U_s/\sqrt{gH}$.  Note that this definition of the Froude number uses the surface speed, $U_s$, rather than the mean flow speed.

\begin{figure}
\includegraphics[width=0.95\textwidth]{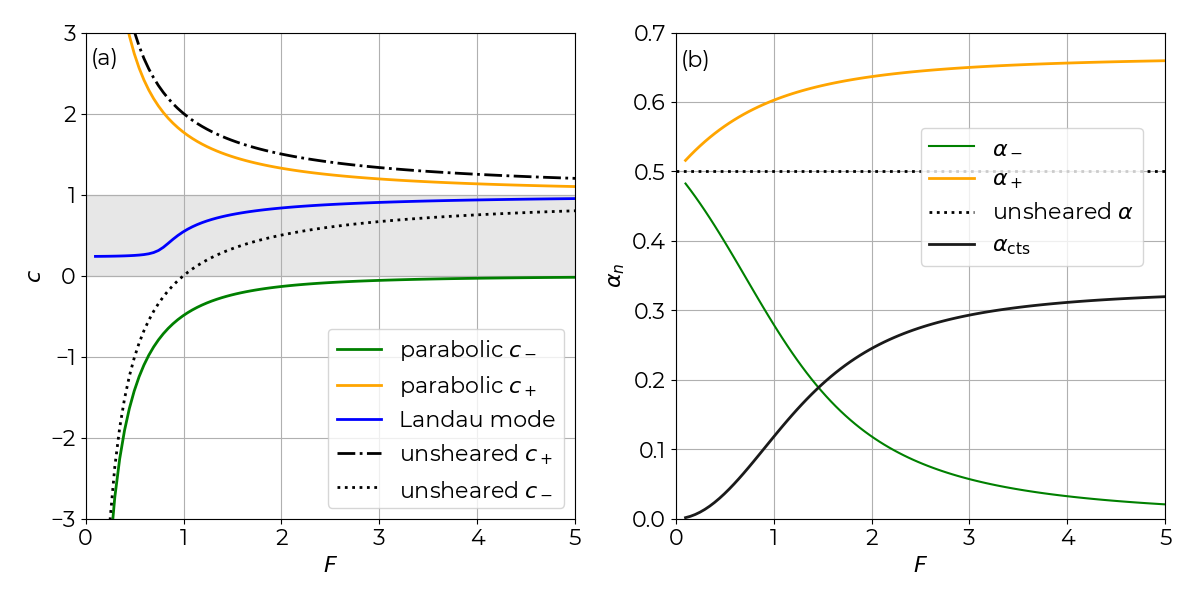} 
\caption{(a) Dispersion relation for the unsheared and parabolic profiles.  For both profiles up- and downstream modes are shown, together with the Landau damped mode from the parabolic profile.  For the parabolic profile, the range of the velocity is shown in gray.  (b) Weights of the normal modes ($\alpha_+, \alpha_-$) and the continuous spectrum ($\alpha_\mathrm{cts}$) for both profiles.  The unsheared profile has $\alpha_\pm = 0.5$ and $\alpha_\mathrm{cts} = 0$.  The values of the weights for each profile sum to unity $\alpha_+ + \alpha_- + \alpha_\mathrm{cts} = 1$.}
\label{f:dispersion}
\end{figure}

Following the analysis above, it is simple to derive the dispersion function $\mathcal{D}(c) = (1 - c)^{-2} - F^{2}$ and to verify that the singularities of $\mathcal{H}(c)$ consist of two poles given by zeros of the dispersion relation $\mathcal{D}(c) = 0$ at $c_\pm = 1 \pm F^{-1}$, with the $\pm$ indicating up- ($-$) and downstream ($+$) propagation (Fig.~\ref{f:dispersion}).  The Laplace inversion results in two discrete wave modes
with amplitudes of the up- and downstream modes equal at $\alpha_\pm = 1/2$, independent of $F$.  No continuous spectrum is present in this case, and the amplitude of the excitation response is partitioned equally between up- and downstream modes.  The total weight of the two modes sums to unity, accounting for the entire response.

\textit{Continuous spectrum solution}.  The second term in (\ref{eq:eta_kt}) represents the continuous spectrum contribution, found by integration around any branch cut singularities in $\mathcal{H}(c)$.  The integration contour, $B$, is thus taken along both sides of the branch cut in the positive sense.  This branch cut contribution appears for velocity profiles with a region of non-zero curvature (or equivalently, vertically varying shear), $U''(z) \neq 0$.  To see this, we make the change of variables $u = U(z)$ in (\ref{eq:D}) to write the dispersion function (in dimensional units) as
\be \label{eq:hyper}
\mathcal{D}(c) = \int_0^{U_s} \frac{\Phi(u)}{(u - c)^2} \: \md u - g^{-1}
\ee 
where we have defined $\Phi(u) \equiv 1/U'(u)$, with the shear, $U'(z)$, written as a function of $u$ via the inversion $z = U^{-1}(u)$.  Here we take $U(z)$ to be a monotonic function, as is typical of channel flows \footnote{This assumption is not necessary, and non-monotonic profiles can be included by splitting the integral in (12) up into the sum of piecewise segments that are monotonic and have a well defined inversion $z = U^{-1}(u)$.}.  Despite the integral in (\ref{eq:hyper}) being undefined along the real $c$ axis, it is known to have well defined limits as the axis is approached from either side.  The limiting values of the integral are given by the Sokhotski-Plemelj-Fox Theorem \cite{fox1957,gala2016}
\be \label{eq:spf_theorem}
\mathcal{D}^\pm(c_r) = \pm i \pi \Phi'(c_r) + \fint_0^{U_s} \frac{\Phi(u)}{(u-c_r)^2} \: \md u - g^{-1}
\ee
where the slash through the integral represents a Hadamard finite part integral \cite{ang2013} that is well defined along the real $c$ axis, represented by $c_r$, and where the plus/minus signs refer to $\mathcal{D}(c_r \pm i\epsilon)$ as $\epsilon \to 0^+$.  It can be verified that if $U''(z) \neq 0$ in some region of $z_1 \leq z \leq z_2$, then $\Phi'(u) \neq 0$ in the corresponding range of $U_1 \leq u \leq U_2$ with $U_i \equiv U(z_i)$, and a branch cut singularity is present.  This range of $U(z)$ that has $U''(z) \neq 0$ will henceforth be referred to as the ``curved range of $U(z)$''.  

This motivates defining the function
\be
\beta(y) \equiv \mathrm{rect}\Big( \frac{y - \bar{U}}{U_2 - U_1} \Big) \: \mathrm{Im}\{ \mathcal{H}^-(y) - \mathcal{H}^+(y) \}
\ee
with $\bar{U} \equiv (U_1 + U_2)/2$, and the rectangular function $\mathrm{rect}(x) \equiv 1$ for $|x|< 1/2$ and zero otherwise.  Here we have defined $\mathcal{H}^\pm$ analogously to $\mathcal{D}^\pm$.  With the above definition, the continuous spectrum part of the solution contained in the integral around the branch cut in (\ref{eq:eta_kt}), denoted $\tilde{\eta}_\mathrm{cts}(k,t)$, can be 
identified with a Fourier transform of $\beta$ in the variable $kt$ [i.e., $\tilde{\beta}(kt)$].  Therefore, in taking the inverse Fourier transform of $\tilde{\eta}_\mathrm{cts}(k,t)$ to get the response of the surface displacement in time and space, $\eta_\mathrm{cts}(x,t)$, we can express the solution for the continuous spectrum as
\be \label{eq:cts_soln}
\eta_\mathrm{cts}(x,t) = \frac{1}{2\pi t} \beta\Big( \frac{x}{t} \Big)
\ee
upon using the scaling property of the Fourier transform $\mathcal{F}^{-1}\{ \tilde{\beta}(kt)\} = t^{-1}\beta(x/t)$.  This analytical solution for the continuous spectrum to an impulsive forcing of the free surface will now be examined in a particular case.


\textit{Waves on a parabolic current}.  In contrast to the constant velocity case above, we now allow for a curvature of the velocity profile [$U''(z) \neq 0$] by taking it to be parabolic: $U(z) = 1 - (1-z)^2$.  The dispersion function can be found analytically through an integration via (\ref{eq:D}) to be (in non-dimensional form)
\be \label{eq:D_parabolic}
\mathcal{D}(c) = -\frac{1}{4(1-c)} \Big[ \frac{1}{\sqrt{1-c}}\log\Big( \frac{\sqrt{1-c} - 1}{\sqrt{1-c} + 1} \Big) + \frac{2}{c} \Big] - F^2 .
\ee
This function has two zeros that lie outside the curved range of $U(z)$ [i.e.~for $0 \leq U(z) \leq 1$] which appear as poles of $\mathcal{H}$, and represent the normal mode solutions.  The locations of these poles as $F$ is varied are shown in Fig.~\ref{f:dispersion}(a), and agree with the results of Burns \cite{burn1953}.  At low $F$, these modes closely resemble the speed of gravity waves in an unsheared flow.  However, as the speed of the upstream mode ($c_-$) approaches the curved range of $U(z)$ the propagation speed asymptotes to its lower boundary, in contrast to the unsheared case.  As the upstream mode deviates from the unsheared case, it experiences a decrease in amplitude, as seen in the decrease of $\alpha_-$ in Fig.~\ref{f:dispersion}(b).  This decrease is accompanied by increases in the amplitude of both the downstream mode ($\alpha_+$) as well as the continuous spectrum. 

\begin{figure}
\includegraphics[width=0.95\textwidth]{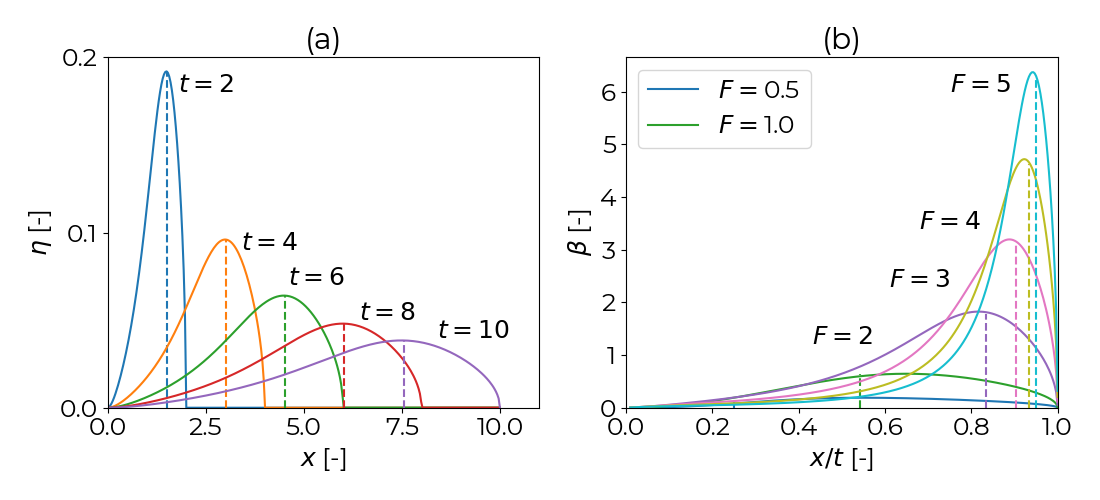} 
\caption{(a) Space-time evolution of the continuous spectrum component of the surface displacement, $\eta_\mathrm{cts}(x,t)$, for the parabolic profile with delta function forcing and $F = 1.5$. (b) The self-similar form for the continuous spectrum evolution of the free surface, $\beta(x/t;F)$, as a function of $F$. The vertical dashed lines in (a,b) correspond to the position and speed, respectively, of the Landau damped mode.}
\label{f:cts_spectrum_xt}
\end{figure}

Taking the principle branches of the logarithm and square root functions, $\mathcal{D}(c)$ and the closely related $\mathcal{H}(c)$ have a branch cut on $0 \leq \mathrm{Re}(c) \leq 1$, within the curved range of $U$.  This results in the presence of a continuous spectrum in the response of this shear flow to a surface excitation that can be quantified using the solution in (\ref{eq:cts_soln}).  The form of these solutions is shown in Fig.~\ref{f:cts_spectrum_xt}.  It represents a surface displacement that is spread out in time by the advection of (the curved range of) $U$, with a maximum amplitude that decreases like $t^{-1}$.  Note that this self-similar evolution for the continuous spectrum component, and its decay by $t^{-1}$, is such that the total vertical displacement, obtained through integrating $\eta_\mathrm{cts}(x,t)$ over all $x$, remains constant.  In other words, the vertical displacement of the continuous spectrum response is spread (or ``dispersed'') in the horizontal by the shear, and decreases in amplitude over time such that the total volume (per unit width) of the response remains fixed.

It is this volume integral, $\alpha_\mathrm{cts} \equiv \int \eta_\mathrm{cts}(x,t) \: \md x$, that is independent of time and plotted along side the weights of the normal modes $\alpha_\pm$, in Fig.~\ref{f:dispersion}(b).  The sum of each of these components in the solution is unity for the given unit delta function forcing, i.e.~$\alpha_+ + \alpha_- + \alpha_\mathrm{cts} = 1$.  Therefore, the continuous spectrum contains a significant fraction of the total excitation response (up to roughly a third at large $F$).  On the other hand, the upstream mode loses its amplitude to both the continuous spectrum and the downstream mode as $F$ increases, so that it plays a vanishing role in carrying the response signal.

\begin{figure}
\includegraphics[width=0.90\textwidth]{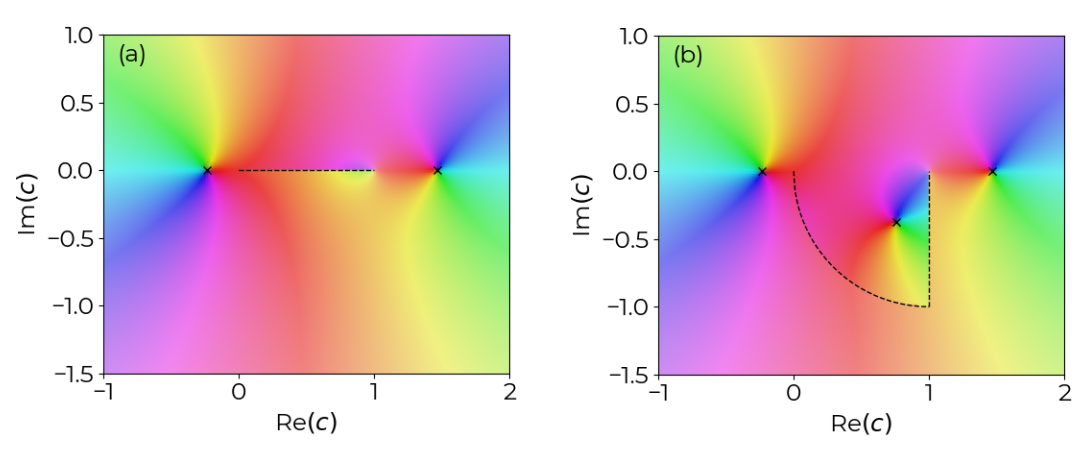} 
\caption{Possible branch cuts of the function $\mathcal{H}(c)$ for $F = 1.5$.  Colors denote the argument of the complex function $\mathcal{H}(c)$, and brightness denotes the amplitude.  (a) Branch cut corresponding to the principle branches of the logarithm and square root functions in $\mathcal{D}(c)$ of equation (\ref{eq:D_parabolic}).  The cut corresponds with $\mathrm{Re}(c)$ in the range of the velocity profile.  (b) Alternate branch cut for $\mathcal{D}(c)$ with cuts corresponding to the negative imaginary axis for the logarithm and square root functions.  Crosses denote the locations of poles, and the dashed lines denote the branch cuts of $\mathcal{H}(c)$.  The damped Landau pole is revealed by the choice of cut in (b). }
\label{f:branch_cuts}
\end{figure}

\textit{Landau damped modes}.  
A continuity between the gravity wave response in the curved range of $U$, and the classical unsheared wave modes can be obtained through a shifting of the branch cut below the real $c$ axis \cite{brig1970}.  This can be done by choosing branch cuts for the logarithmic and square root functions in (\ref{eq:D_parabolic}) along the negative imaginary axis.  The result of this shift is shown in Fig.~\ref{f:branch_cuts}.  As the cut is moved off the real axis, the presence of a pole in $\mathcal{H}(c)$ is revealed on a different branch of the Riemann surface with a negative imaginary part.  This leads to a damped mode that is analogous to the Landau damping effect \cite{land1946,brig1970}.  This Landau damped mode, exhibits a propagation speed that resembles the upstream mode of the unsheared gravity wave dispersion relation for $F \gtrsim 1$, with speed within the curved range of $U$ (Fig.~\ref{f:dispersion}a).  For $F \lesssim 1$ the Landau damped mode resembles the propagation of a long vorticity wave \cite{caul1994,bain1995,carp2010b,book}, which is independent of $F$ (see Appendix).  The propagation speed of the Landau damped mode can also be seen to coincide closely to the peak surface displacement of the continuous spectrum in Fig.~\ref{f:cts_spectrum_xt}.  The Landau damped mode thus provides a link between the classical dispersion relations and the continuous spectrum that describes signal propagation inside the curved range of $U$.

\textit{Summary}.  The continuous spectrum is found to form an important contribution to the response of long sheared gravity waves to excitation.  This response takes the form of three different signal carriers: the up- and downstream normal modes, which follow the dispersion relation, as well as the continuous spectrum describing the response evolution for phase speeds in the curved range of the velocity profile.  At large Froude numbers greater than unity, the upstream mode contributes little to the response, with the bulk contribution by the downstream mode and continuous spectrum, indicating that little communication occurs upstream.  The Landau damped mode facilitates a connection between the dispersion relation and the continuous spectrum that reveals behavior resembling the unsheared case of wave propagation in the curved range of the velocity profile; it fills the spectral gap and provides clarity on the lack of modal solutions in this range of phase speed.  Thus, the behavior of the continuous spectrum around the curved range of the velocity profile is connected to the piecewise and unsheared approximations and is revealed by considering the initial value problem approach.

\begin{acknowledgments}
\textit{Acknowledgments.}  Funding provided by the Helmholtz Association, through project T2 (Ocean Surface Layer Energetics) of the Collaborative Research Center TRR181 ``Energy Transfers in Atmosphere and Ocean'' funded by the Deutsche Forschungsgemeinschaft (DFG) project number 274762653.  Additional support is acknowledged from the European Research Council (ERC) under the European Union's Horizon Europe ERC consolidator grant no.~101088555, `Feedbacks on extreme strorms by ocean turbulent mixing' (FOXSTORM).
\end{acknowledgments}

\appendix

\begin{figure}
\includegraphics[width=0.90\textwidth]{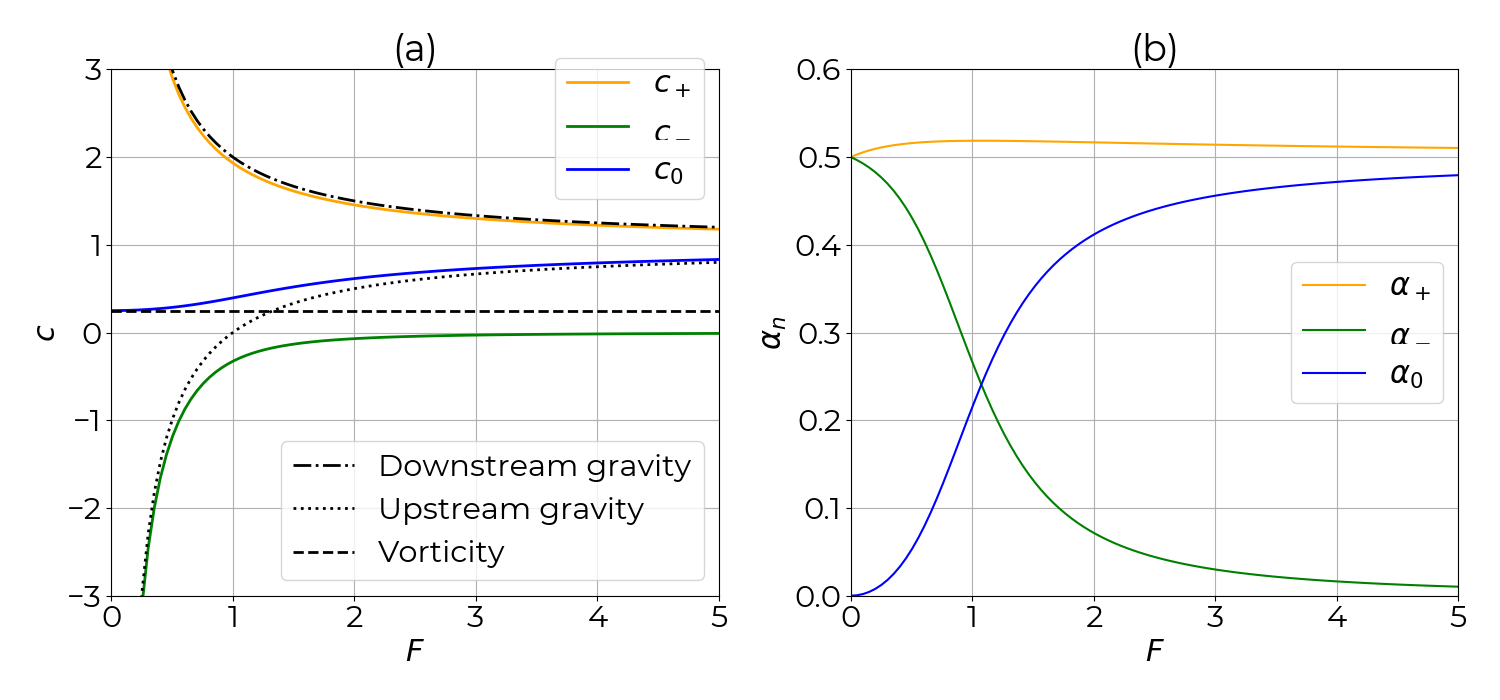} 
\caption{(a) Normal mode phase speeds for the kinked piecewise $U(z)$ profile (solid colored lines).  Plotted additionally are the unsheared gravity wave modes, and the long wave vorticity mode which is independent of $F$.  (b)  Weights of each mode of the kinked profile obtained from the residues with colors matching the modes in panel (a).}
\label{f:kinked_profile}
\end{figure}

\textit{Appendix}.  In addition to the up- and downstream gravity wave modes, shallow water shear flows that exhibit vertical vorticity gradients (i.e.~$U''(z) \neq 0$) are expected to support vorticity wave modes.  These are easily identified in piecewise profiles of $U(z)$ that have kinks at discrete levels where the profile vorticity, $U'(z)$, jumps \footnote{Carpenter and Guha \cite{carp2019} also identified vorticity waves in smooth profiles when the phase speed lies in a region where $U''(z) = 0$.}.  The following profile, non-dimensionalized by the usual scales $U_s$ and $H$, is examined
\be
U(z) = \left\{ \begin{array}{c@{\:}l} 
1 & \textrm{, $h_\ast < z < 1$} \\\relax
z/h_\ast  & \textrm{, $ 0< z \leq h_\ast$} \\   \end{array} \right. ,
\ee
to provide a reference case for comparison to the Landau damped mode of the parabolic profile.  The dispersion relation for this profile can be derived using (\ref{eq:D},\ref{eq:spf_theorem}), to give
\be
c^3 - 2c^2 + (1 - F^{-2})c + h_\ast F^{-2} = 0
\ee
and has 3 zeros associated with the two gravity modes and vorticity mode.  Figure \ref{f:kinked_profile}(a) shows the behavior of the modes for variable $F$ and fixed $h_\ast = 0.25$.  The gravity wave modes of the unsheared velocity profile are plotted as in Figure \ref{f:dispersion}, together with the long vorticity wave speed (dashed line) assuming a solid upper boundary with vanishing vertical velocity.  This vorticity wave speed is found to be $c = \Delta Q H_\mathrm{eff} + U_s$ in dimensional units, with $\Delta Q$ the jump in vorticity across the kink, and $H_\mathrm{eff} \equiv h(H-h)/H$ for dimensional lower layer thickness $h$.  In nondimensional units, the long vorticity wave speed reduces to $c_\ast = h_\ast$, independent of $F$.  Figure \ref{f:kinked_profile}(a) clearly allows for an identification of each mode into gravity and vorticity waves, with a transition in identity between the vorticity and upstream gravity mode around $F \approx 1$.  This transition is accompanied by a change in the dominant weights (Fig.~\ref{f:kinked_profile}b) such that a surface excitation is predominantly carried by the gravity wave modes.  Similar behavior is also seen in the Landau damped mode of Figure \ref{f:dispersion}(a), where the vorticity and gravity modes exist within the curved range of $U(z)$.      

\bibliography{/home/jeffcarp/Documents/latex/references/references_short.bib}

\end{document}